\documentclass[12pt]{article}
\usepackage{amsmath,amssymb}
\newcommand{\sect}[1]{\setcounter{equation}{0}\section{#1}\indent}

\renewcommand{\thefootnote}{\fnsymbol{footnote}}
\newcommand{\EQ}{\begin{equation}}
\newcommand{\EN}{\end{equation}}
\newcommand{\bea}{\begin{eqnarray}}
\newcommand{\ena}{\end{eqnarray}}
\newcommand{\vs}[1]{\vspace{#1 mm}}

\newcommand{\e}{\epsilon}

\newcommand{\uda}{\nearrow \kern-1em \searrow}

\newcommand{\cb}{{\cal B}_0}
\newcommand{\cbh}{\hat{{\cal B}_0}}
\newcommand{\uh}{\hat{U}}
\newcommand{\tphi}{\tilde{\phi}}
\newcommand{\tz}{\tilde{z}}
\newcommand{\tw}{\tilde{w}}

\newcommand{\cl}{{{\cal L}_0}}
\newcommand{\clh}{\hat{\cal L}_0}
\newcommand{\tx}{{\tilde x}}
\newcommand{\ty}{{\tilde y}}
\newcommand{\pbf}{\frac{\pi}{4}}

\newcommand{\la}{\lambda}

\newcommand{\beq}{\begin{equation}} 
\newcommand{\eeq}{\end{equation}}
\newcommand{\beqs}{\begin{eqnarray}} 
\newcommand{\eeqs}{\end{eqnarray}}

\newcommand{\bra}[1]{\langle #1 |}
\newcommand{\ket}[1]{|#1 \rangle}
\def\v{\vert}
\def\r{\rangle}
\def\l{\langle}
\def\p{\partial}

\def\pr{\prime}

\def\nonum{\nonumber}

\def\dstyle{\displaystyle}
\def\t{\tilde}

\def\e{\epsilon}
\def\cd{\cdot}
\def\la{\lambda}
\def\cal{\mathcal}
\def\h{\hat}
\begin{document}

\begin{titlepage}
\setcounter{page}{0}
\begin{flushright}
EPHOU 06-003\\
September 2006\\
\end{flushright}

\vs{6}
\begin{center}
{\bf{{\Large Open String Amplitudes in Various Gauges}}}

\vspace{1cm}
{\large Hiroyuki Fuji, Shinsaku Nakayama and Hisao Suzuki }
\vspace{1cm}

{\em Department of Physics, \\
Hokkaido
University \\  Sapporo, Hokkaido 060 Japan\\
}
\end{center}
\vs{6}

\centerline {{\bf {Abstract}}}
Recently, Schnabl constructed the analytic solution of the open string
 tachyon. Subsequently, the absence of the physical states at the vacuum
 was proved.  The development relies heavily on the use of the gauge
 condition different from the ordinary one. It was shown that the choice
 of gauge simplifies the analysis drastically.  When we perform the
 calculation of the amplitudes in Schnabl gauge, we find that the
 off-shell amplitudes of the Schnabl gauge is still very complicated.  In
 this paper, we propose the use of the propagator in the modified
 Schnabl gauge and show that this modified use of the Schnabl gauge
 simplifies the computation of the off-shell amplitudes drastically. We
 also compute the amplitudes of open superstring in this gauge.
\newpage
\renewcommand{\thefootnote}{\arabic{footnote}}
\setcounter{footnote}{0}
\end{titlepage}
\sect{Introduction}
One of the recent striking achievements in string field theory is the
analytic proof of Sen's conjectures \cite{senconj1,senconj2}. In
\cite{Schnabl}, Schnabl constructed an analytic 
solution for the equation of motion
\bea
Q_B\Phi+\Phi*\Phi=0, \label{equation1}
\ena
in Witten's cubic string field theory \cite{Witten-SFT}
and proved that the height of the tachyon potential at the vacuum is
related to the tension of the D-brane \cite{senconj1}. The consistency
of the solutions has been checked in \cite {Okawa2006,Fuchs2006}.
Subsequently, the Sen's third conjecture which states that there is no
physical state at this vacuum was proved analytically \cite{Ellwood}.

The equation of motion (\ref{equation1}) is a highly non-linear equation
with an infinite numbers of degrees of freedom. In the Siegel gauge
$b_0\Phi=0$, which is traditionally used for the most of the computation,
the equation can be solved by tedious numerical calculation such as
level truncation \cite{KS,SZ,TZ}. In this gauge, the calculations of the
amplitudes are also formidable task \cite{Giddings,Samuel}.

Recent developments of the string field theory rely heavily on the use
of  the proper gauge for the calculation. Schnabl realized that the
gluing rule of string field theory does not match with the Siegel gauge
and used another gauge 
which is more useful
in the star operation \cite{Schnabl}. Subsequent proof of the absence of
the physical degree of freedom also relies heavily on the use of this
gauge \cite{Ellwood}. In \cite{RZ}, the technique has been generalized to 
obtain solutions for a ghost number zero string field equation.

Another problem in the string field theory is the complicated expression
of off-shell amplitudes \cite{Giddings,Samuel}. Recent developments suggest that this gauge simplifies the analysis of the amplitudes. However, the propagator in this gauge turns out  not to be convenient for explicit calculation which was stated in \cite{Schnabl}. 

 In this paper, we propose the use of another gauge for the quantum
 fluctuation fields.  We will show that this choice of gauge for
 the propagator drastically simplifies the calculation of the off-shell
 amplitudes in Witten's cubic string field theory. We also show that
 the modified use can also be applied to WZW-like action
 \cite{N.Berkovits} of open superstring field theory.
 
 This paper is organized as follows. In the next section, we will review
 the star calculus in $\tz$ coordinate and how to calculate four point
 amplitudes. In section 3, we will compute the expression of
 four-tachion off-shell amplitudes in the Schnabl gauge. We will find that
 the amplitudes is rather complicated although it reduces to the known
 amplitudes at on-shell.  In section 4,  We will propose the use of the
 modified Schnabl gauge for the propagator. We will show that the
 modified use of the gauge simplifies the amplitudes drastically. We
 will give a formula for four point amplitudes in this gauge. It will be
 shown that four point amplitudes of non-Abelian gauge theories reduce
 the quartic terms which are expected from the Yang-Mills action
 \cite{BS}. In section 5, we show how to use the modified gauge in WZW-like
 action of the superstring to compute the off-shell string
 amplitudes. 
In section 6, we compute the four point 
amplitude for tachyons in GSO($-$) sector and the effective quartic 
terms of the gauge fields in the zero momentum limit. 
 The final section is devoted to some discussions.
 \sect{Witten's cubic interaction in $\tz$ coordinate and amplitudes}
In Witten's open string field theory, the gluing condition simplifies
 in the coordinates $\tz=\arctan z$. In this coordinate, the primary
 state $\phi(z)$ of dimension $h$ is given by \cite{Schnabl}
\bea
\tphi (\tz)=(\frac{dz}{d\tz})^h\phi(\tan\tz)=(\cos\tz)^{-2h}\phi(\tan\tz).
\ena
The scaling generator can be written by the energy momentum tensor in this coordinate as
\bea
\cl=\oint\frac{d\tz}{2\pi i}\tz T_{\tz \tz}(\tz)=L_0+\sum_{k=0}^\infty \frac{2(-1)^{k+1}}{4k^2-1}L_{2k}.
\ena
The scaling operator $U_r$ can be defined as
\bea
U_r=(\frac{2}{r})^\cl.
\ena
The action of $U_r$ with the field of conformal dimension $h$ is simply given by
\bea
U_r\tphi(\tz)U_r^{-1}=(\frac{2}{r})^h\tphi(\frac{2}{r}\tz).\label{scaling}
\ena
A non-trivial property of this operator is 
\bea
e^{-\beta \clh}=U_{2+2\beta}^\dagger U_{2+2\beta},
\ena
where 
\bea
\clh=\cl+\cl^\dagger.
\ena
Because of this property, it seems convenient to define an operator
\bea
{\uh}_r=U_r^\dag U_r.
\ena
We can easily find  $\uh_2=1$ and the product rule is
\bea
\uh_r\uh_s=\uh_{r+s-2}.
\ena

 Three vertex of string field theory defines a mapping gluing two fields into one state. In the coordinate system, the mapping can be simply given by
 \bea
 \tphi_1(0)\ket{0}* \tphi_2(0)\ket{0}=\uh_3\tphi_1(\frac{\pi}{4})\tphi_2(-\frac{\pi}{4})\ket{0}.
 \ena
 Since the BPZ conjugate of the states are given by $z'=-1/z$, the conjugation can be expressed as $\tz'=\tz\pm \pi/2$ in $\tz$ coordinate. For example, the conjugate of the above state is given by 
 \bea
(\uh_3\tphi_1(\frac{\pi}{4})\tphi_2(-\frac{\pi}{4})\ket{0})^\dagger= \bra{0}\tphi_2(-\frac{\pi}{4}-\frac{\pi}{2})\tphi_1(\frac{\pi}{4}+\frac{\pi}{2})\uh_3.
 \ena
More generally, the gluing of the states of the form $\uh \tphi$ takes a simple form
\bea
\uh_r{\tphi_1}(\tx)\ket{0}*\uh_s{\tphi_2}(\ty)\ket{0}=\uh_{r+s-1}\tphi_1(\tx+\pbf(s-1)){\tphi}_2(\ty-\pbf(r-1))\ket{0}.
\ena
In order to obtain the exact solutions, Schnabl used a gauge \cite{Schnabl}
\begin{eqnarray}
\cb\Phi=0, 
\label{Schnabl gauge}
\end{eqnarray}
where $\cb$ is the zero mode of the $b$ ghost in the $\tz$ coordinate
\bea
\cb=\oint \frac{d\tz}{2\pi i}\tz b(\tz)=b_0+\sum_{k=0}^\infty \frac{2(-1)^{k+1}}{4k^2-1}b_{2k}.
\ena
Its anti-commutator with BRST charge is given by $\{Q_B,\cb \}=\cl$. For later convenience, we define 
\bea
\cbh=\cb+\cb^\dag,
\ena
from which we find a relation $\{Q_B,\cbh\}=\clh$.
 The advantage of the Schnabl gauge is that the form of the fields
 including ghosts and $\clh, \cbh$ close under the star operations and
 the action of the BRST charge \cite{Schnabl}. This choice turns out to
 be crucial for obtaining exact solution of eq. (\ref{equation1}).
 
Useful identities including $\cbh$ are
\bea
\cbh \uh_r&=&\uh_r \cbh,\\
\cbh U_r^\dag&=&\frac{2}{r}U_r^\dag \cbh,\label{commutator1}\\
\cbh(\phi_1*\phi_2)&=&\frac{\pi}{2}(-1)^{{\rm{gh}}(\phi_1)}(B_1\phi_1)*\phi_2+\phi_1*(\cbh\phi_2),
\ena
where $B_1=b_1+b_{-1}$. We find that anti-commutators
\bea
\{\cb,{\tilde c}(\tz)\}&=&\tz,\nonumber\\
\{B_1,{\tilde c}(\tz)\}&=&1,
\ena
are also useful for the calculation of the amplitudes.
Correlation functions of the fields in the $\tz$ coordinates are 
summarized as
\bea
\langle\partial{\tilde X}^\mu(\tz)\partial{\tilde X}^\nu(\tw)
\rangle &=& -\frac{\alpha'}{2}\eta^{\mu\nu}\frac{1}{\sin^2(\tz-\tw)},
\nonumber\\
\langle\t{c}(\t{x})\t{c}(\t{y})\t{c}(\t{w})\rangle
&=&\sin(\t{x}-\t{y})\sin(\t{y}-\t{w})\sin(\t{w}-\t{x}).
\ena
In this section, we will show how the above relations will be used to
obtain the amplitudes.

Witten's cubic action is given by
\bea
S=-\frac{1}{g^2}\left[ \frac{1}{2}\langle\Phi, Q_B\Phi\rangle
+\frac{1}{3}\langle\Phi*\Phi*\Phi\rangle\right].
\ena
To find the effective action, we will use the background field method \cite{BS}. We separate the field into the background field $\phi_b$ and quantum fluctuation $R$ as
\bea
\Phi=\phi_b+R.
\ena
We consider the path integral of the field $R$. The contribution of $R$ to the action is 
\bea
S=-\frac{1}{g^2}\left[ \frac{1}{2}\langle R,
Q_BR\rangle+\langle\phi_b*\phi_b*R\rangle+\langle R*\phi_b*R\rangle
+\frac{1}{3}\langle R*R*R\rangle\right].
\ena
To find the effective action, we need to consider the propagator of the
quantum fluctuation field $R$ and fix the gauge.
We can obtain the four point process by shifting the field $R$ \cite{BS}
\bea
R \rightarrow R-{\cal P}\phi_b*\phi_b,
\ena
where ${\cal P}$ is the propagator which satisfies a relation
\bea
Q_B{\cal P}=1.
\ena
As a result of the shift of the quantum fluctuation field $R$,
we find that the quartic interaction term is given in $\tz$ coordinates as
\bea
A_4=\frac{1}{2}\bra{0}\tphi_b(-\pbf-\frac{\pi}{2})\tphi_b(\pbf+\frac{\pi}{2})\uh_3,{\cal P}\uh_3\tphi_b(\pbf)\tphi_b(-\pbf)\ket{0}. \label{a4}
\ena
Since the propagator ${\cal P}$ depends on the gauge of the field $R$,
the four point off-shell amplitudes also depend on the gauge condition. 
In the following, we shall show that the dependence of the gauge choice vanishes at on-shell.
\sect{Four point amplitudes in the Schnabl gauge}
We are now going to compute the amplitudes in the Schnabl gauge. Let us
consider the fields of the form $\phi_b =c(z)V_1(z)$.
For example, the tachyon and photon vertices are simply given by 
$\phi_b =c(z)e^{ikX(z)}$ and $\phi_b =\e_\mu(k)c(z)\sqrt{\frac{2}{\alpha^{\pr}}}\partial X^\mu(z)e^{ikX(z)}$. 

 In the Schnabl gauge (\ref{Schnabl gauge}), the propagator is given by
 \bea
 {\cal P}=\frac{\cb}{\cl}Q_B\frac{\cb^\dag}{\cl^\dag}. 
 \ena
For the computation of the amplitude in this gauge, we need two
 Schwinger parameters for this propagator
 \bea
 {\cal P}&=&\mathcal{B}_{0}\int^{\infty}_{0}dt_{1}e^{-t_{1}\mathcal{L}_{0}}Q_{B}\mathcal{B}_{0}^{\dagger}
	      \int^{\infty}_{0}dt_{2}e^{-t_{2}\mathcal{L}_{0}^{\dagger}}\nonum\\
	   &=&\int^{\infty}_{0}dt_{1}\int^{\infty}_{0}dt_{2}
	      \mathcal{B}_{0}U_{T_{1}}Q_{B}\mathcal{B}_{0}^{\dagger}U_{T_{2}}^{\dagger},
 \ena
where $T_{1}=2e^{t_{1}},\,T_{2}=2e^{t_{2}}$.

 Using the commutation rules, we find
 \begin{eqnarray}
 	\hat{U}_{3}\cal{B}_{0}U_{T_{1}}Q_{B}\cal{B}_{0}^{\dagger}U_{T_{2}}^{\dagger}\hat{U}_{3}
	=\hat{U}_{3}\cal{B}_{0}U_{T_{1}}\hat{U}_{3}-\hat{U}_{3}\cal{B}_{0}U_{T_{1}}\cal{B}_{0}^{\dagger}U_{T_{2}}^{\dagger}
	\hat{U}_{3}Q_{B}.
 \end{eqnarray}
Thus in the Schnabl gauge, the four point amplitude $\eqref{a4}$ is written as follows.
 \begin{eqnarray}
 	&&A_{4}=\frac{1}{2}\int^{\infty}_{0}dt\bra{0}\tphi_{1}(\pbf+\frac{\pi}{2})
  	\tphi_{2}(-\pbf-\frac{\pi}{2})\hat{U}_{3}\cal{B}_{0}U_{T}\hat{U}_{3}\tphi_{3}(\pbf)\tphi_{4}(-\pbf)\ket{0}\nonum\\
      &&\quad-\frac{1}{2}\int^{\infty}_{0}dt_{1}\int^{\infty}_{0}dt_{2}\bra{0}\tphi_{1}(\pbf+\frac{\pi}{2})
  	\tphi_{2}(-\pbf-\frac{\pi}{2})\hat{U}_{3}\cal{B}_{0}U_{T_{1}}\cal{B}_{0}^{\dagger}U_{T_{2}}^{\dagger}
	\hat{U}_{3}Q_{B}\tphi_{3}(\pbf)\tphi_{4}(-\pbf)\ket{0}.\nonum\\ \label{schnabl_amp}
 \end{eqnarray}
 In the case that all $\phi_{i}\,'s$ are primary field with weight $h_{i}$, using
 \begin{eqnarray}
 	\hat{U}_{3}\cal{B}_{0}U_{T}\hat{U}_{3}=U_{\frac{3T+2}{T}}^{\dagger}\left(\frac{3T}{3T+2}\cal{B}_{0}-
	\frac{2}{3T+2}\cal{B}_{0}^{\dagger}\right)U_{\frac{3T+2}{2}},
 \end{eqnarray}
and eq. \eqref{scaling}, the correlator in the first term can be simplified as
\begin{eqnarray}
&&\Biggl\langle
\tilde{\phi}_1\left(\frac{\pi}{4}+\frac{\pi}{2}\right)
\tilde{\phi}_2\left(-\frac{\pi}{4}-\frac{\pi}{2}\right)
U_3^{\dagger}U_3
{\cal B}_0U_{T_1}
U_3^{\dagger}U_3
\tilde{\phi}_3\left(\frac{\pi}{4}\right)
\tilde{\phi}_4\left(-\frac{\pi}{4}\right)
\Biggr\rangle
\nonumber \\
&=&\Biggl\langle
\tilde{\phi}_1\left(\t{z}_{1}\right)
\tilde{\phi}_2\left(-\t{z}_{1}\right)
\left(\frac{3T_1}{3T_1+2}{\cal B}_0-\frac{2}{3T_1+2}{\cal B}_0^{\dagger}\right)
\nonumber 
\tilde{\phi}_3\left(\t{z}_{2}\right)
\tilde{\phi}_4\left(-\t{z}_{2}\right)\Biggr\rangle\\
&&\times\left(\frac{2T_1}{3T_1+2}\right)^{h_1+h_2}
\left(\frac{4}{3T_1+2}\right)^{h_3+h_4},
\end{eqnarray}
where
 \begin{eqnarray}
 	\t{z}_{1}=\frac{\pi(2T_{1}+1)}{3T+2},\quad \t{z}_{2}=\frac{\pi}{3T_{1}+2}.
 \end{eqnarray}

As an example, let us consider the four point amplitude of tachyons.
Substituting into the tachyon vertex operator, the first term of eq. $\eqref{schnabl_amp}$ yields
\begin{eqnarray}
 	&&\frac{1}{2}(2\pi)^{26}\delta^{26}(\textstyle\sum k_{i})
      \dstyle\int_{0}^{1/2}dy\;y^{-2-\alpha^{\prime}s}
	(1-y)^{-2-\alpha^{\prime}u} 
	\left(\frac{2T_1}{3T_1+2}\right)^{\alpha^{\prime}(k_1^2+k_2^2)-2}
	\left(\frac{4}{3T_1+2}\right)^{\alpha^{\prime}(k_3^2+k_4^2)-2} \nonum\\
	&&\quad\times
	\sin(2\tilde{z}_1)^{2-\alpha^{\prime}(k_1^2+k_2^2)}
	\sin(2\tilde{z}_2)^{2-\alpha^{\prime}(k_3^2+k_4^2)}
	\bigl(\sin(\tilde{z}_1+\tilde{z}_2)\bigr)
	^{4-\alpha^{\prime}(k_1^2+k_2^2+k_3^2+k_4^2)} \nonum\\
	&&\quad\times\bigl(\sin(\tilde{z}_1-\tilde{z}_2)\bigr)
	^{4-\alpha^{\prime}(k_1^2+k_2^2+k_3^2+k_4^2)},
 \end{eqnarray}
where new variable $y$ is introduced such as
 \begin{eqnarray}
 	&&y=-\frac{\sin(2\tilde{z}_1)\sin(2\tilde{z}_2)}
{\sin^2(\tilde{z}_1-\tilde{z}_2)},
\quad
1-y=\frac{\sin^2(\tilde{z}_1+\tilde{z}_2)}
{\sin^2(\tilde{z}_1-\tilde{z}_2)},
 \nonum\\
&&
y=1/2\quad ({\rm for}\;t=0),
\quad y=0\quad ({\rm for}\;t=\infty).
\end{eqnarray}

The second term of $\eqref{schnabl_amp}$ which vanishes for on-shell amplitude is rather complicated.
Using
 \begin{eqnarray}
 	U_{3}^{\dagger}U_{3}\mathcal{B}_{0}U_{T_{1}}\mathcal{B}_{0}^{\dagger}U_{T_2}^{\dagger}U^{\dagger}_{3}U_{3}
	=-\frac{4}{3T_{1}+3T_{2}-4}U^{\dagger}_{\frac{3T_{1}+3T_{2}-4}{T_{1}}}\mathcal{B}_{0}^{\dagger}
	\mathcal{B}_{0}U_{\frac{3T_{2}+3T_{1}-4}{T_{2}}},
 \end{eqnarray}
and evaluating correlator, we get
 \begin{eqnarray}
 	&&-\frac{1}{2}\int^{\infty}_{0}dt_{1}\int^{\infty}_{0}dt_{2}
      \frac{4(-\alpha^{\pr}k_{3}^{2}-\alpha^{\pr}k_{4}^{2}+2)}{3T_{1}+3T_{2}-4}
	\left(\frac{2}{V}\right)^{-2+\alpha^{\pr}(k_{1}^{2}+k_{2}^{2})}
	\left(\frac{2}{W}\right)^{-2+\alpha^{\pr}(k_{3}^{2}+k_{4}^{2})}
	 \nonum\\
	&&\times \dstyle (2\pi)^{26}\delta^{26}(\textstyle\sum k_{i}) |\sin\frac{\pi}{V}|^{2\alpha^{\pr}k_{1}\cdot k_{2}}
	|\sin\frac{\pi}{W}|^{2\alpha^{\pr}k_{3}\cdot k_{4}}
	|\cos(\frac{\pi}{2V}-\frac{\pi}{2W})|^{2\alpha^{\pr}(k_{1}\cdot k_{3}+k_{2}\cdot k_{4})}
	\nonum\\
	&&\times|\cos(\frac{\pi}{2V}+\frac{\pi}{2W})|^{2\alpha^{\pr}(k_{1}\cdot k_{4}+k_{2}\cdot k_{3})}
	\frac{\pi}{2V}\Bigl(\cos(\frac{\pi}{V})+\cos(\frac{\pi}{W})\Bigr)
	\Bigl(\frac{\pi}{W}\cos(\frac{\pi}{W})-\sin(\frac{\pi}{W})\Bigr), \nonum\\
 \end{eqnarray}
where
 \begin{eqnarray}
 	\frac{2}{V}=\frac{2T_{1}}{3(T_{1}+T_{2})-4}
	,\quad\frac{2}{W}=\frac{2T_{2}}{3(T_{1}+T_{2})-4}.
 \end{eqnarray}

   \sect{Four point amplitudes of tachyons and gauge fields in the modified Schnabl gauge}
   In the previous section, we have seen that the four point amplitudes
   in the Schnabl gauge is very complicated. The complication stems from the
   form of the propagator in this gauge. To avoid these difficulties, we
   will use the propagator in the gauge $\cbh R=0$ not in the Schnabl gauge
   $\cb R=0$ for the quantum fluctuation field.  In this gauge, the propagator can be written as
   \bea
   {\cal P}=\frac{\cbh}{\clh},
   \ena
   which is manifestly self-conjugate and commutes with $\uh$ which appears in the amplitudes.
   
   The first question is whether the choice of gauge conditions may influence the physical observables. We will argue that on-shell amplitudes does not depend on the choice of the gauge conditions.
         
   Suppose we have modified $b_0$ as
   \bea
   b_0'=b_0+\sum_{n=-\infty,n\neq 0}^{\infty}a_n b_n
   \ena
    with some parameters $a_n$. We use a gauge for the state $b_0'R=0$, the propagator of this gauge is given by
   \bea
   {\cal P}'=\frac{b_0'}{L_0'},
   \ena
   where
   \bea
   L_0'=\{Q_B,b_0'\}=L_0+\sum_{n=-\infty,n\neq 0}^{\infty}a_n L_n.
   \ena
   We will compute the variation of the propagator with respect to $a_n$;
   \bea
   \frac{\partial {\cal P}}{\partial a_n}=b_n\frac{1}{L_0'}-b_0'\frac{1}{L_0'}L_n\frac{1}{L_0'}.
   \ena
   Because of the relations $L_0'=\{Q_B,b_0'\}$ and $L_n=\{Q_B,b_n\}$, this expression can be rewritten as
   \bea
   \frac{\partial {\cal P}}{\partial a_n}=\{Q_B,b_0'\frac{1}{L_0'}b_n\frac{1}{L_0'}\}.\label{brsexact}
   \ena
    Eq. (\ref{brsexact}) implies that the dependence on the parameters is
   BRST closed and decouples from the correlation functions of BRST
   exact states. This statement may be an extension of the propagator of
   the usual gauge fields. 
That is to say, 
even though the propagator of gauge fields contains the gauge parameters, the total amplitudes do not depend on the parameters.
    
   Since the off-shell states themselves are not BRST exact states, the
   off-shell amplitudes depend on the gauge conditions. However, the
   final physical quantities obtained by the use of the off-shell
   amplitudes should be BRST exact. Therefore, it is expected that the
   choice of gauge for the internal states will not affect the physical observables.

Let us compute the off-shell amplitudes in the modified Schnabl gauge. We find that the quartic interaction term is given by
\bea
A_4=\frac{1}{2}\bra{0}\tphi_1(-\pbf-\frac{\pi}{2})\tphi_2(\pbf+\frac{\pi}{2})\uh_3,\frac{\cbh}{\clh}\uh_3\tphi_3(\pbf)\tphi_4(-\pbf)\ket{0}.
\ena
We will use Schwinger parametrization as
\bea
\frac{\cbh}{\clh}=\int_0^\infty d\beta \cbh e^{-\beta\clh}=\int_0^\infty d\beta\cbh\uh_{2\beta+2}.
\ena
Using the manipulation rule $\uh_r\uh_s=\uh_{r+s-2}$ and the fact that $\cbh$ and $\uh_r$ commute with each other, we easily find
\bea
A_4=\frac{1}{2}\int_0^\infty d\beta\bra{0}\tphi_1(-\pbf-\frac{\pi}{2})\tphi_2(\pbf+\frac{\pi}{2})\cbh\uh_{2\beta+4}\tphi_3(\pbf)\tphi_4(-\pbf)\ket{0}.
\ena
By expressing $\uh_{2\beta+4}=U^\dagger_{2\beta+4}U_{2\beta+4}$ and relations (\ref{scaling}) and (\ref{commutator1}),  we can move $U^\dagger_{2\beta+4}$ to the left and $U_{2\beta+4}$ to the right to find
\bea
A_4
&=&\frac{1}{2}\int_0^\infty \frac{d\beta}{(\beta+2)^{\sum_i h_i+1}}\bra{0}\tphi_1(-\pbf(\frac{1}{\beta+2})-\frac{\pi}{2})\tphi_2(\pbf(\frac{1}{\beta+2})+\frac{\pi}{2})\nonumber\\
&{}&\qquad\times \cbh\tphi_3(\pbf(\frac{1}{\beta+2}))\tphi_4(-\pbf(\frac{1}{\beta+2}))\ket{0}.
\ena
By changing variable by $t=1/(\beta+2)$, we arrive at the following
formula of four point amplitudes; 
\begin{eqnarray}
A_4=\frac{1}{2}\int_0^{1/2}dt\; {t^{\sum_ih_i-1}}\bra{0}\tphi_1(-\pbf
t-\frac{\pi}{2})\tphi_2(\pbf t+\frac{\pi}{2})(\cb+\cb^\dag)\tphi_3(\pbf
t)\tphi_4(-\pbf t)\ket{0}.
\label{A_4}
\end{eqnarray}
Let us apply this formula to the four point tachyon amplitudes. 
The tachyon vertex operators are $\phi_{i}=ce^{ik_{i}\cdot X}$, and the following correlator included in the formula (4.11) is
easily evaluated by using eqs. (2.18) and (2.19)
 \begin{eqnarray}
  	&&\l 0\v \t{c}e^{ik_{1}\cdot \t{X}}(\frac{\pi}{4}t+\frac{\pi}{2})\t{c}e^{ik_{2}\cdot \t{X}}(-\frac{\pi}{4}t-\frac{\pi}{2})
      (\mathcal{B}_{0}+\mathcal{B}_{0}^{\dagger})\t{c}e^{ik_{3}\cdot \t{X}}(\frac{\pi}{4}t)
      \t{c}e^{ik_{4}\cdot \t{X}}(-\frac{\pi}{4}t)\v 0\r \nonum\\
      &&\quad=(2\pi)^{26}\delta^{26}
	(\textstyle{\sum_{i}}k_{i})\dstyle
	\pi t\sin(\frac{\pi}{2}t)^{2\alpha^{\pr}(k_{1}\cdot k_{2}+k_{3}\cdot k_{4})+1}
      \cos(\frac{\pi}{2}t)^{2\alpha^{\pr}(k_{1}\cdot k_{4}+k_{2}\cdot k_{3})+1}.
 \end{eqnarray}
Changing variable $y=\sin^{2}\frac{\pi}{2}t$, we find the four tachyon amplitude as
 \begin{eqnarray}
 	A_{4}&=&\frac{1}{2}(2\pi)^{26}\delta^{26}(\textstyle{\sum_{i}}k_{i})\dstyle\int^{1/2}_{0}dy\,
	t(y)^{\alpha^{\pr}\sum k_{i}^{2}-4}
	y^{-\alpha^{\pr}s-\alpha^{\pr}\sum_{i}k_{i}^{2}/2}
	(1-y)^{-\alpha^{\pr}u-\alpha^{\pr}\sum_{i}k_{i}^{2}/2},\nonum\\
 \end{eqnarray}
where $t(y)=\frac{2}{\pi}\arcsin \sqrt{y}$.
Note that the integral appearing in the amplitudes are very simple although we cannot get any analytic expression for generic values of momenta.

To get the full off-shell tachyon amplitude, we must consider all $4!$ permutations of external momenta $k_{i}\;(i=1,2,3,4)$.
The sum of these 4! permutations give six different terms, and each of these has a factor 4
 \begin{eqnarray}
 	A_{4}&=&2(2\pi)^{26}\delta^{26}(\textstyle{\sum_{i}}k_{i})\dstyle
      [I(s,u)+I(u,s)+I(u,t)+I(t,u)+I(t,s)+I(s,t)], \nonum\\
 \end{eqnarray}
where
 \begin{eqnarray}
 	I(s,u)=\int^{1/2}_{0}dy\;t(y)^{\alpha^{\pr}\sum k_{i}^{2}-4}
	y^{-\alpha^{\pr}s-\alpha^{\pr}\sum_{i}k_{i}^{2}/2}(1-y)^{-\alpha^{\pr}u-\alpha^{\pr}\sum_{i}k_{i}^{2}/2}.
 \end{eqnarray}

Let us consider the case of on-shell amplitudes where $\alpha'k_i^2=1$. Using this on-shell condition, $I(s,u)$ becomes
 \begin{eqnarray}
 	I(s,u)=\int^{1/2}_{0}dy\;y^{-\alpha^{\pr}s-2}(1-y)^{-\alpha^{\pr}u-2}.
 \end{eqnarray}
Therefore, $(s\leftrightarrow u)$ term contributes to the range 
 $\frac{1}{2}<y<1$ after changing variable $y\rightarrow 1-y$, so that
 \begin{eqnarray}
 	I(s,u)+I(u,s)=\int^{1}_{0}dy\;y^{-\alpha^{\pr}s-2}(1-y)^{-\alpha^{\pr}u-2}=B(-1-\alpha^{\pr}s, -1-\alpha^{\pr}u),
 \end{eqnarray}
where $B(a,b)$ is the Euler beta function
 \begin{eqnarray}
 	B(a,b)=\int^{1}_{0}dy\;y^{a-1}(1-y)^{b-1}.
 \end{eqnarray}
Other four terms can be combined in the same way. 
Totally we have the following well-known expression for the tachyon amplitudes;
\begin{eqnarray}
 	A_{4}=2(2\pi)^{26}\delta^{26}
	(\textstyle{\sum_{i}}k_{i})\dstyle[B(-\alpha(s),-\alpha(u))+B(-\alpha(u),-\alpha(t))+B(-\alpha(t),-\alpha(s))],
\end{eqnarray}
where $\alpha(s)=1+\alpha^{\pr}s$.

In the case of non-Abelian gauge fields, the procedures are quite similar.
However, even in the off-shell amplitude, we have to impose the transversality condition $\e_{i}\cdot k_{i}=0$
on the vector vertex operators $\phi_{i}=\e_{\mu\,i}c\p X^{\mu}e^{ik_{i}\cdot X}\;(i=1,2,3,4)$, since the formula 
(4.11) is applicable only to the primary operators.

The off-shell four vector amplitude is 
 \begin{eqnarray}
 	A_{4}&=&\frac{1}{2}(2\pi)^{26}\delta^{26}(\textstyle{\sum_{i}}k_{i})\dstyle
	\Biggl[J(s,u)\mathrm{Tr}\,(\la^{a_{1}}\la^{a_{2}}\la^{a_{3}}\la^{a_{4}})
	+J(u,s)\mathrm{Tr}\,(\la^{a_{1}}\la^{a_{4}}\la^{a_{3}}\la^{a_{2}}) \nonum\\
	&&+J(u,t)\mathrm{Tr}\,(\la^{a_{1}}\la^{a_{3}}\la^{a_{2}}\la^{a_{4}})
	+J(t,u)\mathrm{Tr}\,(\la^{a_{1}}\la^{a_{4}}\la^{a_{2}}\la^{a_{3}}) \nonum\\
	&&+J(t,s)\mathrm{Tr}\,(\la^{a_{1}}\la^{a_{2}}\la^{a_{4}}\la^{a_{3}})
	+J(s,t)\mathrm{Tr}\,(\la^{a_{1}}\la^{a_{3}}\la^{a_{4}}\la^{a_{2}})
	\Biggr]\cal{F}(y;\e,k),\label{gaugeamplitudes}
 \end{eqnarray}
where
 \begin{eqnarray}
 	J(s,u)=\int^{1/2}_{0}dy\;t(y)^{\alpha^{\pr}\sum k_{i}^{2}}
	y^{-\alpha^{\pr}s-\alpha^{\pr}\sum_{i}k_{i}^{2}/2}(1-y)^{-\alpha^{\pr}u-\alpha^{\pr}\sum_{i}k_{i}^{2}/2},
 \end{eqnarray}
and the explicit form of function $\cal{F}(y;\e,k)$ which includes 
polarization vectors $\e_{i}\;(i=1,2,3,4)$ 
is given in Appendix A.

Imposing on-shell conditions $k_{i}^{2}=0$, 
we find 
  \begin{eqnarray}
 	&&\hspace{-0.6cm}A_{4}=(2\pi)^{26}\delta^{26}(\textstyle{\sum_{i}}k_{i})
      \dstyle\int_{0}^{1} dy\;
	\Biggl[y^{-\alpha^{\pr}s}(1-y)^{-\alpha^{\pr}u}\,\mathrm{Tr}\,(\la^{a_{1}}\la^{a_{2}}\la^{a_{3}}\la^{a_{4}}
	+\la^{a_{1}}\la^{a_{4}}\la^{a_{3}}\la^{a_{2}}) \nonum\\
	&&+y^{-\alpha^{\pr}u}(1-y)^{-\alpha^{\pr}t}\,\mathrm{Tr}\,(\la^{a_{1}}\la^{a_{3}}\la^{a_{2}}\la^{a_{4}}+
	\la^{a_{1}}\la^{a_{4}}\la^{a_{2}}\la^{a_{3}}) \nonum\\
	&&+y^{-\alpha^{\pr}t}(1-y)^{-\alpha^{\pr}s}\mathrm{Tr}(\la^{a_{1}}\la^{a_{2}}\la^{a_{4}}\la^{a_{3}}+
	\la^{a_{1}}\la^{a_{3}}\la^{a_{4}}\la^{a_{2}})\Biggr]\cal{F}(y;\e,k).
 \end{eqnarray}
When we take the limit $k\rightarrow 0$, this expression reduces to the
four point interactions required for Yang-Mills action \cite{BS}.
\sect{Effective quartic interaction for open superstring}
In the previous section, we have shown how the modified use of the Schnabl
gauge simplifies the computation of open string amplitudes. In this 
section, we will extend the analysis to the open superstrings. We are
going to  derive the formula of the effective quartic coupling for the
superstring using WZW-like action \cite{N.Berkovits}
 \begin{eqnarray}
 	S=\frac{1}{4g^{2}}\Bigl\l (e^{-\Phi}Q_{B}e^{\Phi})(e^{-\Phi}\eta_{0}e^{\Phi})
	  -\int_{0}^{1}dt(e^{-t\Phi}\p_{t}e^{t\Phi}\{(e^{-t\Phi}Q_{B}e^{t\Phi})
	  ,(e^{-t\Phi}\eta_{0}e^{t\Phi})\}\Bigr\r. \label{BPS}
 \end{eqnarray}
The ordinary choice of the gauge is
 \begin{eqnarray}
 	b_{0}\Phi=0,\quad \xi_{0}\Phi=0. \label{bxigauge}
 \end{eqnarray}
The cubic terms in this action are extracted as 
 \begin{eqnarray}
 	S_{3}=\frac{1}{6g^{2}}\Bigl[\l (Q_{B}\Phi)\Phi(\eta_{0}\Phi)\r-\l (Q_{B}\Phi )(\eta_{0}\Phi)\Phi\r\Bigr].
 \end{eqnarray}
Expanding the field around the background $(\Phi\rightarrow \Phi+R)$, 
we get the terms linear in $R$
 \begin{eqnarray}
 	&&\l (Q_{B}\Phi)R(\eta_{0}\Phi)\r-\l (Q_{B}\Phi )(\eta_{0}\Phi)R\r+\l (Q_{B}R)\Phi(\eta_{0}\Phi)\r-\l (Q_{B}R )(\eta_{0}\Phi)\Phi\r \nonum\\
	&&+\l (Q_{B}\Phi)\Phi(\eta_{0}R)\r-\l (Q_{B}\Phi )(\eta_{0}R)\Phi\r
\nonumber \\
&&	
=3\Bigl[\l (Q_{B}\Phi)R(\eta_{0}\Phi)\r-\l (Q_{B}\Phi )(\eta_{0}\Phi)R\r\Bigr].
 \end{eqnarray}
Therefore, the total action is
 \begin{eqnarray}
 	S&=&-\frac{1}{2g^{2}}\l\eta_{0} R,Q_{B}R\r
          -\frac{1}{2g^{2}}\l R,(Q_{B}\Phi)*(\eta_{0}\Phi)+(\eta_{0}\Phi)*(Q_{B}\Phi)\r+\cdots \nonum\\
	 &=&-\frac{1}{2g^{2}}\l\eta_{0} R,Q_{B}R\r
	    +\frac{1}{2g^{2}}\l \eta_{0}R,\xi_{0}\Bigl\{(Q_{B}\Phi) *(\eta_{0}\Phi)+(\eta_{0}\Phi)*(Q_{B}\Phi)
	    \Bigr\}\r+\cdots,
\nonumber \\
&&
 \end{eqnarray}
where we have used $\xi_{0}R=0$.
Shifting the quantum fluctuation field $R$ by
 \begin{eqnarray}
 	R\rightarrow R-\frac{1}{2}\frac{b_{0}}{L_{0}}\xi_{0}
      \Bigl\{(Q_{B}\Phi) *(\eta_{0}\Phi)+(\eta_{0}\Phi)*(Q_{B}\Phi)\Bigr\} 
 \end{eqnarray}
to eliminate the terms linear in R, we get the effective quartic coupling
 \begin{eqnarray}
 	S^{(4)}&=&-\frac{1}{2g^{2}}\frac{1}{4}\l Q_{B}\frac{b_{0}}{L_{0}}\xi_{0}\Phi^{(2)}
                ,\eta_{0}\frac{b_{0}}{L_{0}}\xi_{0}\Phi^{(2)}\r \nonum\\
	   	 &=&-\frac{1}{2g^{2}}\frac{1}{4}\l\Phi^{(2)},\frac{b_{0}}{L_{0}}\xi_{0}\Phi^{(2)}\r, \label{su4pt}
 \end{eqnarray}
where
 \begin{eqnarray}
 	\Phi^{(2)}=(Q_{B}\Phi)*(\eta_{0}\Phi)+(\eta_{0}\Phi)*(Q_{B}\Phi). \label{phi2}
 \end{eqnarray}
\par
In particular, when the on-shell condition $\eta_{0}Q_{B}\Phi=0$ is satisfied, we can rewrite $\Phi^{(2)}$ as 
 \begin{eqnarray*}
 	\Phi^{(2)}=-\eta_{0}\Bigl\{(Q_{B}\Phi)*\Phi-\Phi*(Q_{B}\Phi)\Bigr\}.
 \end{eqnarray*}
 Therefore, the effective quartic coupling is given by \cite{BS}
 \begin{eqnarray}
 	S^{(4)}=-\frac{1}{2g^{2}}\frac{1}{4}\l Q_{B}\frac{b_{0}}{L_{0}}(\Phi*Q_{B}\Phi-Q_{B}\Phi*\Phi)
                ,\eta_{0}\frac{b_{0}}{L_{0}}(\Phi*Q_{B}\Phi-Q_{B}\Phi*\Phi)\r. \label{on-shell}
 \end{eqnarray}
\par
Again we are going to use $\t{z}$ coordinates in the same way as in Witten's cubic action. 
Gauge conditions $\eqref{bxigauge}$ are transformed by $U_{\tan}$ into
 \begin{eqnarray}
 	\cal{B}_{0}\t{\Phi}=0,\quad \t{\xi}_{0}\t{\Phi}=0.
 \end{eqnarray}
Therefore, states which satisfy eq. $\eqref{bxigauge}$ automatically satisfy these conditions in $\t{z}$ coordinates.
The first one is what we call the Schnabl gauge condition.
We will next consider the modified Schnabl gauge. We define the modified Schnabl gauge for the superstring as
 \begin{eqnarray}
 	\hat{\cal{B}}_{0}R=0,\quad \t{\xi}_{0}R=0.
 \end{eqnarray}
Imposing this modified Schnabl gauge for the fluctuation $R$, 
the effective quartic term $\eqref{su4pt}$ is rewritten as
 \begin{eqnarray}
 	S^{(4)}=-\frac{1}{2g^{2}}\frac{1}{4}\l(Q_{B}\t{\Phi})*(\eta_{0}\t{\Phi})+(\eta_{0}\t{\Phi})*(Q_{B}\t{\Phi})
		    ,\frac{\hat{\cal{B}}_{0}}{\hat{\cal{L}}_{0}}\t{\xi}_{0}\Bigl\{
		    (Q_{B}\t{\Phi})*(\eta_{0}\t{\Phi})+(\eta_{0}\t{\Phi})*(Q_{B}\t{\Phi})\Bigr\}\r. \nonum\\ \label{su4pt_sch} 
 \end{eqnarray}
\sect{Four point amplitude of tachyons and gauge fields}
We are now going to compute the off-shell four point amplitude of tachyons in the modified Schnabl gauge.
In order to deal with GSO($-$) sector that tachyons appear, we consider the string field action for the non-BPS D-brane:
 \begin{eqnarray}
 	S=\frac{1}{4g^{2}}\Bigl\l (e^{-\h{\Phi}}\h{Q}_{B}e^{\h{\Phi}})(e^{-\h{\Phi}}\h{\eta}_{0}e^{\h{\Phi}})
	  -\int_{0}^{1}dt(e^{-t\h{\Phi}}\p_{t}e^{t\h{\Phi}}\{(e^{-t\h{\Phi}}\h{Q}_{B}e^{t\h{\Phi}})
	  ,(e^{-t\h{\Phi}}\h{\eta}_{0}e^{t\h{\Phi}})\}\Bigr\r, \label{non-BPSac}
 \end{eqnarray}
In this action, $2\times 2$ internal Chan-Paton factors are added both to the vertex operators and to $Q_{B}$ 
and $\eta_{0}$. 
The tachyon vertex operator is then written as
 \begin{eqnarray}
 	\h{\phi}=\xi ce^{-\phi}e^{ik\cd X}\otimes \sigma_{1}.
 \end{eqnarray}
$Q_{B}$ and $\eta_{0}$ are tensored with $\sigma_{3}$
 \begin{eqnarray}
 	\hat{Q}_{B}=Q_{B}\otimes \sigma_{3},\quad \hat{\eta}_{0}=\eta_{0}\otimes\sigma_{3}.
 \end{eqnarray}
\par
Since the algebraic structure of this non-BPS action is completely identical to that of BPS action $\eqref{BPS}$,
we can get the same formula for the quartic coupling as eq. $\eqref{su4pt_sch}$ up to a factor $2$
which compensate the trace of the internal CP matrices.
Finally, we find the effective quartic couping 
 \begin{eqnarray}
 	S^{(4)}=-\frac{1}{4g^{2}}\frac{1}{4}\l(\h{Q}_{B}\t{\Phi})*(\h{\eta}_{0}\t{\Phi})
                 +(\h{\eta}_{0}\t{\Phi})*(\h{Q}_{B}\t{\Phi})
		    ,\frac{\hat{\cal{B}}_{0}}{\hat{\cal{L}}_{0}}\t{\xi}_{0}\Bigl\{
		    (\h{Q}_{B}\t{\Phi})*(\h{\eta}_{0}\t{\Phi})+(\h{\eta}_{0}\t{\Phi})*(\h{Q}_{B}\t{\Phi})\Bigr\}\r. \nonum\\
 \end{eqnarray}
Corresponding amplitude is given by the same procedure as in the bosonic string
 \begin{eqnarray}
 	 A_{4}&=&-\frac{1}{4g^{2}}\frac{1}{4}\int_{0}^{\infty}d\beta
		    \l\Bigl(\hat{\phi}_{\eta}(-\frac{\pi}{4}-\frac{\pi}{2})\hat{\phi}_{Q}(\frac{\pi}{4}+\frac{\pi}{2})
		    +\hat{\phi}_{Q}(-\frac{\pi}{4}-\frac{\pi}{2})\hat{\phi}_{\eta}(\frac{\pi}{4}+\frac{\pi}{2})\Bigr)
		 \nonum\\
		  &&\times\hat{\cal{B}}_{0}\hat{U}_{2\beta+4}\t{\xi}_{0}
		    \Bigl(\hat{\phi}_{Q}(\frac{\pi}{4})\hat{\phi}_{\eta}(-\frac{\pi}{4})
		    +\hat{\phi}_{\eta}(\frac{\pi}{4})\hat{\phi}_{Q}(-\frac{\pi}{4})\Bigr)\r \nonum\\
		 &=&-\frac{1}{4g^{2}}\frac{1}{4}\int_{0}^{1/2}dt\;t^{\sum h_{i}-1}
		    \l\Bigl(\hat{\phi}_{\eta}(-\frac{\pi}{4}t-\frac{\pi}{2})\hat{\phi}_{Q}(\frac{\pi}{4}t+\frac{\pi}{2})
		    +\hat{\phi}_{Q}(-\frac{\pi}{4}t-\frac{\pi}{2})\hat{\phi}_{\eta}(\frac{\pi}{4}t+\frac{\pi}{2})\Bigr) \nonum\\
		  &&\times\hat{\cal{B}}_{0}\t{\xi}_{0}
		    \Bigl(\hat{\phi}_{Q}(\frac{\pi}{4}t)\hat{\phi}_{\eta}(-\frac{\pi}{4}t)
		    +\hat{\phi}_{\eta}(\frac{\pi}{4}t)\hat{\phi}_{Q}(-\frac{\pi}{4}t)\Bigr)\r, \label{su_tachyon}
 \end{eqnarray} 
where we have defined 
 \begin{eqnarray}
 	\hat{\phi}_{Q}&=&(Q_{B}\otimes \sigma_{3})(\t{\phi}\otimes\sigma_{1}) \nonum\\
	              &=&\Bigl[(-\alpha^{\pr}k^{2}+\frac{1}{2})\p\t{c}\t{c}\t{\xi}e^{-\t{\phi}}e^{ik\cd \t{X}}
	            +\sqrt{2\alpha^{\pr}}k^{\mu}\t{\psi}_{\mu}\t{c}e^{ik\cd\t{X}}-\t{\eta} e^{\t{\phi}}e^{ik\cd \t{X}}
	\Bigr]\otimes i\sigma_{2}, \\
      \hat{\phi}_{\eta}&=&(\eta_{0}\otimes \sigma_{3})(\t{\phi}\otimes\sigma_{1}) \nonum\\
	                 &=&\t{c}e^{-\t{\phi}}e^{ik\cd \t{X}}
	\otimes i\sigma_{2}.
 \end{eqnarray}
Evaluating all correlation functions in eq. $\eqref{su_tachyon}$, we get
 \begin{eqnarray}
 	 A_{4}&=&-\frac{1}{4g^{2}}\frac{2}{4}\int_{0}^{1/2}dt\;t^{\sum (\alpha^{\pr}k_{i}^{2}-\frac{1}{2})-1}
		    (2\pi)^{d}\delta^{d}(\textstyle\sum k_{i})\dstyle
	      y^{-\alpha^{\pr}s-\alpha^{\pr}\sum k_{i}^{2}/2}(1-y)^{-\alpha^{\pr}t-\alpha^{\pr}\sum k_{i}^{2}/2}
	      \nonum\\
	     &&\hspace{-0.5cm}\times\Biggl[\eta_{\mu\nu}\pi t\sin\frac{\pi}{2}t\cos\frac{\pi}{2}t
	     \Biggl(2\alpha^{\pr}(k_{1}^{\mu}k_{3}^{\nu}
	       +k_{2}^{\mu}k_{4}^{\nu})-2\alpha^{\pr}(k_{1}^{\mu}k_{4}^{\nu}+k_{2}^{\mu}k_{3}^{\nu})
	     \frac{1}{\cos^{2}\frac{\pi}{2}t}\Biggr) \nonum\\
 	     &&\hspace{-0.5cm}+(-\alpha^{\pr}k_{1}^{2}+\frac{1}{2})
		\Biggl(\frac{1}{2}\cot\frac{\pi}{4}(\pi t\cos\pi t-\sin\pi t)
	     +\frac{1}{2}\cot\frac{\pi}{4}\sec\frac{\pi}{2}(\pi t-\sin\pi t)\Biggr) \nonum\\
	     &&\hspace{-0.5cm}-(-\alpha^{\pr}k_{3}^{2}+\frac{1}{2})
	     \Biggl(\frac{1}{8}\pi t(4\cos\pi t-\sin\pi t)\tan\frac{\pi t}{4}
	      -\frac{1}{8}\pi t\sec\frac{\pi t}{2}(4+\sin\pi t)\tan\frac{\pi t}{4}\Biggr) \nonum\\
	     &&\hspace{-0.5cm}
	     -(-\alpha^{\pr}k_{4}^{2}+\frac{1}{2})\Biggl(
	     \frac{1}{8}\pi t\sec\frac{\pi t}{2}(-4+\sin\pi t)\tan\frac{\pi t}{4}
	     +\frac{1}{8}\pi t(4\cos\pi t+\sin\pi t)\tan\frac{\pi t}{4}\Biggr) \nonum\\
	     &&\hspace{-0.5cm}+(-\alpha^{\pr}k_{2}^{2}+\frac{1}{2})\Biggl(
	     \frac{1}{2}\cot\frac{\pi}{4}\sec\frac{\pi}{2}(\pi t\cos\pi t-\sin\pi t)
	     +\frac{1}{2}\cot\frac{\pi}{4}(\pi t\cos\pi t-\sin\pi t)\Biggr)
	     \Biggr].\nonum \\
 \end{eqnarray}
It was shown that the WZW-like action reproduces the on-shell four point amplitudes correctly \cite{N.Berkovits and C.T.Echevarria}.
Here we will see the consistency of the above off-shell amplitude at on-shell.  
Imposing $\alpha^{\pr}k_{i}^{2}=1/2\;(i=1,2,3,4)$, we find that on-shell four tachyon amplitude is obtained as
 \begin{eqnarray}
 	 A_{4}&=&\frac{-1}{4g^{2}}(2\pi)^{d}\delta^{d}(\textstyle\sum k_{i})\dstyle\int_{0}^{1/2}dy
	      \Biggl[-y^{-\alpha^{\pr}s-1}(1-y)^{-\alpha^{\pr}u-2}(\alpha^{\pr}u+1) \nonum\\
	      &&+y^{-\alpha^{\pr}s-1}(1-y)^{-\alpha^{\pr}u-1}(\alpha^{\pr}t+1)\Biggr].\label{sutachyon}
 \end{eqnarray}
We rewrite the first term of eq. $\eqref{sutachyon}$
by partial integration
 \begin{eqnarray}
 	\int_{0}^{1/2}dy\;y^{-\alpha^{\pr}s-1}(1-y)^{-\alpha^{\pr}u-2}(\alpha^{\pr}u+1)
      &=&\int_{0}^{1/2}dy\;y^{-\alpha^{\pr}s-1}\p_{y} (1-y)^{-\alpha^{\pr}u-1} \nonum\\
	&=&(1/2)^{-\alpha^{\pr}s-1}(1/2)^{-\alpha^{\pr}u-1} \nonum\\
	&&+(\alpha^{\pr}s+1)\int_{0}^{1/2}dy\;y^{-\alpha^{\pr}s-2}(1-y)^{-\alpha^{\pr}u-1}.\nonum\\
 \end{eqnarray}
Summation over $4!$ permutations of the momenta yields to Euler beta functions. Using identity 
 \begin{eqnarray*}
	-(-\alpha^{\pr}s-1)B(-\alpha^{\pr}s-1,-\alpha^{\pr}u)
 	&=&-(-\alpha^{\pr}s-1)\frac{\Gamma(-\alpha^{\pr}s-1)\Gamma(-\alpha^{\pr}u)}{\Gamma(-\alpha^{\pr}s-\alpha^{\pr}u-1)}\\
 	&=&\frac{-\Gamma(-\alpha^{\pr}s)\Gamma(-\alpha^{\pr}u)}{\Gamma(-\alpha^{\pr}s-\alpha^{\pr}u)}
 	 (-\alpha^{\pr}s-\alpha^{\pr}u-1)\\
	&=&B(-\alpha^{\pr}s,-\alpha^{\pr}u)(\alpha^{\pr}t+1),
 \end{eqnarray*}
we find 
 \begin{eqnarray}
 	  	A_{4}&=&\frac{-1}{g^{2}}(2\pi)^{d}\delta^{d}(\textstyle\sum k_{i})\dstyle
		\Biggl[2(1+\alpha^{\pr}u)B(-\alpha^{\pr}s,-\alpha^{\pr}t)
		-(1/2)^{-\alpha^{\pr}s-1}(1/2)^{-\alpha^{\pr}t-1} \nonum\\
	     &&+2(1+\alpha^{\pr}t)B(-\alpha^{\pr}s,-\alpha^{\pr}u)
		-(1/2)^{-\alpha^{\pr}s-1}(1/2)^{-\alpha^{\pr}u-1} \nonum\\
	     &&+2(1+\alpha^{\pr}s)B(-\alpha^{\pr}t,-\alpha^{\pr}u)
		-(1/2)^{-\alpha^{\pr}t-1}(1/2)^{-\alpha^{\pr}u-1}\Biggr]. \label{sutach_tot}
 \end{eqnarray}

In order to get the complete four point amplitude, we have to consider
summation over the permutations of momenta.
In addition, since the super string field action $\eqref{non-BPSac}$ is non-polynomial, quartic 
coupling which describes contact interaction of four string fields
 \begin{eqnarray}
 	S_{4}=\frac{1}{4!\cd 2g^{2}}\Biggl[-2\l (\hat{Q}_{B}\hat{\Phi})\hat{\Phi}(\hat{\eta}_{0}\hat{\Phi})
	\hat{\Phi}\r+\l(\hat{Q}_{B}\hat{\Phi})\hat{\Phi}^{2}(\hat{\eta}_{0}\hat{\Phi})\r
	  +\l (\hat{Q}_{B}\hat{\Phi})(\hat{\eta}_{0}\hat{\Phi})\hat{\Phi}^{2}\r\Biggr], \label{contact}
 \end{eqnarray}
also contributes to the four point amplitudes. 
We expect that the on-shell four point tachyon amplitude in total
agrees with the first quantization result \cite{N.D.Lambert and
I.Sachs}. 
The contributions from the contact interaction $\eqref{contact}$ is given by
 \begin{eqnarray}
 	A_{4}^{\rm contact}&=&-\frac{1}{g^{2}}(2\pi)^{d}\delta^{d}(\textstyle \sum k_{i})\dstyle
	       \Biggl[(1/2)^{-\alpha^{\pr}s-1}(1/2)^{-\alpha^{\pr}u-1}
		+(1/2)^{-\alpha^{\pr}u-1}(1/2)^{-\alpha^{\pr}t-1} \nonum\\ 
		&&+(1/2)^{-\alpha^{\pr}t-1}(1/2)^{-\alpha^{\pr}s-1}\Biggr].
 \end{eqnarray}
This contribution just cancels the extra terms in
 eq. $\eqref{sutach_tot}$ and the result agrees with that of the first
 quantization \cite{N.D.Lambert and I.Sachs}
\begin{eqnarray}
A_4+A_4^{\rm contact}
&=&
\frac{-2}{g^{2}}(2\pi)^{d}\delta^{d}(\textstyle\sum k_{i})\dstyle
		\Biggl[(1+\alpha^{\pr}u)B(-\alpha^{\pr}s,-\alpha^{\pr}t)
\nonumber \\
&&
+(1+\alpha^{\pr}t)B(-\alpha^{\pr}s,-\alpha^{\pr}u)
+(1+\alpha^{\pr}s)B(-\alpha^{\pr}t,-\alpha^{\pr}u)
\Biggr].
\end{eqnarray}

Four point amplitude of gauge fields is obtained in the similar way. 
For simplicity, we only consider 
the gauge fields with zero momenta. The vertex operator is given by
 \begin{eqnarray*}
 	\phi=\t{c}\t{\xi} e^{-\t{\phi}}\t{\psi}^{\mu}.
 \end{eqnarray*}
From eq. $\eqref{on-shell}$ on-shell four point amplitude is given by
 \begin{eqnarray}
 	A_{4}&=&-\frac{1}{2g^{2}}\frac{1}{4}\l (\phi*\phi_{Q}-\phi_{Q}*\phi)
                ,\eta_{0}\frac{\h{\cal{B}}_{0}}{\h{\cal{L}}_{0}}(\phi*\phi_{Q}-\phi_{Q}*\phi)\r \nonum\\
	     &=&-\frac{1}{8g^{2}}\int_{0}^{1/2}\frac{dt}{t}
		  \l\biggl(\phi_{Q}(-\frac{\pi}{4}t-\frac{\pi}{2})\phi(\frac{\pi}{4}t+\frac{\pi}{2})
		  -\phi(-\frac{\pi}{4}t-\frac{\pi}{2})\phi_{Q}(\frac{\pi}{4}t+\frac{\pi}{2})\biggr) \nonum\\
	     &&\times\eta_{0}\h{\cal{B}}_{0}\biggl(\phi(\frac{\pi}{4}t)\phi_{Q}(-\frac{\pi}{4}t)
		  -\phi_{Q}(\frac{\pi}{4}t)\phi(-\frac{\pi}{4}t)\biggr)\r,
 \end{eqnarray} 
where
 \begin{eqnarray}
 	\phi_{Q}&=&Q_{B}\phi \nonum\\
	&=&i\sqrt{\frac{2}{\alpha^{\pr}}}\t{c}\p \t{X}^{\mu}.
 \end{eqnarray}
After the evaluation of this correlation functions, integration can be done explicitly
 \begin{eqnarray}
  	A_{4}&=&\frac{1}{8g^{2}}\int_{0}^{1/2}dy\;
		  \Biggl(2\eta^{\nu\sigma}\eta^{\mu\rho}+2\frac{\eta^{\nu\rho}\eta^{\mu\sigma}}{(1-y)^{2}}\Biggr) \nonum\\
	     &=&\frac{1}{4g^{2}}\Biggl(\frac{1}{2}\eta^{\mu\rho}\eta^{\nu\sigma}+\eta^{\mu\sigma}\eta^{\nu\rho}\Biggr).
 \end{eqnarray}
Contribution from eq. $\eqref{contact}$ is 
 \begin{eqnarray}
 	A_{4}^{\rm contact}=\frac{1}{g^{2}}\Biggl(\frac{1}{8}\eta^{\mu\rho}\eta^{\nu\sigma}
	      -\frac{1}{2}\eta^{\mu\sigma}\eta^{\nu\rho}\Biggr).
 \end{eqnarray}
Then the total four point amplitude is just the Yang-Mills quartic coupling.
 \begin{eqnarray*}
 	A_{4}=\frac{1}{g^{2}}\Biggl(\frac{1}{4}\eta^{\mu\rho}\eta^{\nu\sigma}
	      -\frac{1}{4}\eta^{\mu\sigma}\eta^{\nu\rho}\Biggr),
 \end{eqnarray*}
 which was first pointed out in \cite{BS} via the Siegel gauge.
\sect{Discussions}
We have obtained the formula for four point amplitudes in $\tz$ coordinates. 
Even in the Schnabl gauge, the off-shell amplitudes are very complicated. 
In this paper, we proposed the use of the modified version of the Schnabl gauge for intermediate states
 and showed that the most of the calculations can be achieved efficiently in this gauge. 
We also applied this modified gauge for the open superstring field theory.

 More interesting question is whether the Schnabl gauge is effective for
 obtaining closed string amplitudes.  
In the closed string field theory, the calculations of the 
amplitudes are so difficult even at on-shell.
It would be interesting to investigate whether the amplitudes of the closed string fields could be obtained in the Schnabl gauge.

We have postponed the arguments about the physical meaning of modified 
use of the Schnabl gauge in the external fields.
It is not clear whether this condition fixes the gauge uniquely.
At the linearized level, it might be shown that this gauge condition is
 consistent by the method used for the Siegel gauge. 
\\

\noindent{\large{\bf Acknowledegments}}

\noindent{The work of H. F. is supported in part by JSPS Research Fellowships
for Young Scientists.
The work of H. S. is supported in part by the Grant-in-Aid for Scientific
Research No. 16081201 from Ministry of Education, Science, Culture and
Sports of Japan.}

\appendix
\sect{Kinematical factor $\mathcal{F}$}
In this appendix, we list the kinetic factor which appeared in eq. (\ref{gaugeamplitudes}).
 \begin{eqnarray*}
      &&\hspace{-1.5cm}\mathcal{F}(y;\e,k)=
	\e_{1}\!\cdot\!\e_{2}\e_{3}\!\cdot\!\e_{4}y^{-2}
	+\e_{1}\!\cdot\!\e_{3}\e_{2}\!\cdot\!\e_{4}
	+\e_{1}\!\cdot\!\e_{4}\e_{2}\!\cdot\!\e_{3}(1-y)^2\\
	&&-2\alpha^{\pr}\Biggl[\e_{1}\!\cdot\!\e_{2}
	(k_{2}\cdot \e_{3})\Biggl\{(k_{2}\cdot \e_{4})(1-y)^{-1}
	+(k_{3}\cdot \e_{4})y^{-1}(1-y)^{-1}\Biggr\}\\
	&&
	-\e_{1}\!\cdot\!\e_{2}(k_{4}\cdot \e_{3})\Biggr\{
	(k_{2}\cdot \e_{4})y^{-1}
	+(k_{3}\cdot \e_{4})y^{-2}\Biggr\}\\
	&&+\e_{1}\!\cdot\!\e_{3}(k_{3}\cdot \e_{2})
	\Biggl\{(k_{3}\cdot \e_{4})y^{-1}(1-y)^{-1}
	+(k_{2}\cdot \e_{4})(1-y)^{-1}\Biggr\}\\
	&&-\e_{1}\!\cdot\!\e_{3}(k_{4}\cdot \e_{2})\Biggl\{(k_{3}\cdot \e_{4})y^{-1}
	+(k_{2}\cdot \e_{4})\Biggr\}\\
	&&-\e_{1}\!\cdot\!\e_{4}(k_{4}\cdot \e_{3})\Biggl\{(k_{4}\cdot \e_{2})y^{-1}
	+(k_{3}\cdot \e_{2})y^{-1}(1-y)^{-1}\Biggr\}\\
	&&+\e_{1}\!\cdot\!\e_{4}(k_{2}\cdot \e_{3})\Biggl\{(k_{4}\cdot \e_{2})(1-y)^{-1}
	+(k_{3}\cdot \e_{2})(1-y)^{-2}\Biggr\}\\
	&&-\e_{2}\!\cdot\!\e_{3}(k_{2}\cdot \e_{1})\Biggl\{(k_{2}\cdot \e_{4})y^{-1}
	+(k_{1}\cdot \e_{4})y^{-1}(1-y)^{-1}\Biggr\}\\
	&&+\e_{2}\!\cdot\!\e_{3}(k_{4}\cdot \e_{1})\Biggl\{(k_{2}\cdot \e_{4})(1-y)^{-1}
	+(k_{1}\cdot \e_{4})(1-y)^{-2}\Biggr\}\\
	&&+\e_{2}\!\cdot\!\e_{4}
	(k_{2}\cdot \e_{3})\Biggl\{(k_{2}\cdot \e_{1})y^{-1}(1-y)^{-1}
	+(k_{3}\cdot \e_{1})(1-y)^{-1}\Biggr\}\\
	&&+\e_{2}\!\cdot\!\e_{4}(k_{1}\cdot \e_{3})\Biggl\{(k_{2}\cdot \e_{1})y^{-1}
      +(k_{3}\cdot \e_{1})\Biggr\}\\
	&&-\e_{3}\!\cdot\!\e_{4}(k_{3}\cdot \e_{2})
	\Biggl\{(k_{3}\cdot \e_{1})(1-y)^{-1}
	+(k_{2}\cdot \e_{1})y^{-1}(1-y)^{-1}\Biggr\}\\
      &&+\e_{3}\!\cdot\!\e_{4}(k_{1}\cdot \e_{2})\Biggl\{(k_{3}\cdot \e_{1})y^{-1}
	+(k_{2}\cdot \e_{1})y^{-2}\Biggr)\Biggr]\\
	&&+4\alpha^{\pr\;2}\Biggl\{-\e_{1}\cdot k_{3}\e_{2}\cdot k_{4}\e_{3}\cdot k_{2}\e_{4}\cdot k_{2}
	+\e_{1}\cd k_{3}\e_{2}\cd k_{4}\e_{3}\cd k_{4}\e_{4}\cd k_{2}(1-y)y^{-1} \\
	&&+\e_{1}\cd k_{3}\e_{2}\cd k_{3}\e_{3}\cd k_{4}\e_{4}\cd k_{2}y^{-1}
	+\e_{1}\cd k_{4}\e_{2}\cd k_{4}\e_{3}\cd k_{4}\e_{4}\cd k_{2}y^{-1}
 \end{eqnarray*}
\newpage
 \begin{eqnarray}
	&&-\e_{1}\cd k_{3}\e_{2}\cd k_{4}\e_{3}\cd k_{2}\e_{4}\cd k_{3}y^{-1} 
	+\e_{1}\cd k_{3}\e_{2}\cd k_{4}\e_{3}\cd k_{4}\e_{4}\cd k_{3}(1-y)y^{-2} \nonum\\
	&&+\e_{1}\cd k_{3}\e_{2}\cd k_{3}\e_{3}\cd k_{4}\e_{4}\cd k_{3}y^{-2} 
	+\e_{1}\cd k_{4}\e_{2}\cd k_{4}\e_{3}\cd k_{4}\e_{4}\cd k_{3}y^{-2} \nonum\\
	&&-\e_{1}\cd k_{3}\e_{2}\cd k_{3}\e_{3}\cd k_{2}\e_{4}\cd k_{2}(1-y)^{-1} 
 	-\e_{1}\cd k_{4}\e_{2}\cd k_{4}\e_{3}\cd k_{2}\e_{4}\cd k_{2}(1-y)^{-1} \nonum\\
	&&+\e_{1}\cd k_{4}\e_{2}\cd k_{3}\e_{3}\cd k_{4}\e_{4}\cd k_{2}y^{-1}(1-y)^{-1} 
	-\e_{1}\cd k_{3}\e_{2}\cd k_{3}\e_{3}\cd k_{2}\e_{4}\cd k_{3}y^{-1}(1-y)^{-1} \nonum\\
	&&-\e_{1}\cd k_{4}\e_{2}\cd k_{4}\e_{3}\cd k_{2}\e_{4}\cd k_{3}y^{-1}(1-y)^{-1} 
	+\e_{1}\cd k_{4}\e_{2}\cd k_{3}\e_{3}\cd k_{4}\e_{4}\cd k_{3}y^{-2}(1-y)^{-1} \nonum\\
	&&-\e_{1}\cd k_{4}\e_{2}\cd k_{3}\e_{3}\cd k_{2}\e_{4}\cd k_{2}(1-y)^{-2} 
	-\e_{1}\cd k_{4}\e_{2}\cd k_{3}\e_{3}\cd k_{2}\e_{4}\cd k_{3}y^{-1}(1-y)^{-2}\Biggr\}.\nonum\\
 \end{eqnarray}

\end{document}